\documentclass[11pt,sort&compress]{elsarticle}

\makeatletter
  \long\def\pprintMaketitle{\clearpage
  \iflongmktitle\if@twocolumn\let\columnwidth=\textwidth\fi\fi
  \resetTitleCounters
  \def\baselinestretch{1}%
  \printFirstPageNotes
  \begin{center}%
 \thispagestyle{pprintTitle}%
   \def\baselinestretch{1}%
    {\large\bf\@title}\par\vskip5pt
    \normalsize\elsauthors\par\vskip5pt
    \footnotesize\itshape\elsaddress\par\vskip10pt
    \end{center}%
  \vskip6pt
  \ifvoid\absbox\else\unvbox\absbox\par\vskip10pt\fi
  \ifvoid\keybox\else\unvbox\keybox\par\vskip10pt\fi
  \gdef\thefootnote{\arabic{footnote}}%
  }
\makeatother

\newcommand\blfootnote[1]{%
  \begingroup
  \renewcommand\thefootnote{}\footnote{#1}%
  \addtocounter{footnote}{-1}%
  \endgroup
}

\journal{}
\usepackage{amsmath}
\usepackage{amsfonts}
\usepackage{amssymb}
\usepackage{lipsum}
\usepackage{graphicx}
\usepackage{xspace}
\usepackage{epsf}
\usepackage{setspace}
\usepackage{float}
\usepackage{booktabs}
\usepackage{pifont}

\usepackage{siunitx}
\usepackage{iftex}
\usepackage[
    protrusion=true,
    activate={true,nocompatibility},
    final,
    tracking=true,
    factor=1100]{microtype}

\usepackage[margin=1in]{geometry}

\usepackage{stmaryrd}
\SetSymbolFont{stmry}{bold}{U}{stmry}{m}{n}

\usepackage[]{xcolor}
\usepackage{bm}
\newcommand{\ve}[1]{\bm{#1}}

\newcommand{\bu}{\ve{u}}

\newcommand{\bx}{\ve{x}}

\newcommand{\bI}{\ve{I}}

\newcommand{\dd}{\text{d}}
\newcommand{\ddd}{\dd R \dd \dot{R}}

\newcommand{\Rdot}{\dot{R}}
\newcommand{\Rddot}{\ddot{R}}

\newcommand{\vbs}{\vec{\ve{s}}}
\newcommand{\vbF}{\vec{\ve{F}}}
\newcommand{\vbg}{\vec{\ve{g}}}
\newcommand{\vbq}{\vec{\ve{q}}}

\newcommand{\hw}{\widehat{w}}

\newcommand{\hR}{\widehat{R}}
\newcommand{\hRdot}{\widehat{\dot{R}}}

\newcommand{\mom}{\mu}
\newcommand{\bmom}{\boldsymbol{\mu}}
\newcommand{\vbmom}{\vec{\bmom}}

\definecolor{lightblue}{rgb}{0.63, 0.74, 0.78}
\definecolor{seagreen}{rgb}{0.18, 0.42, 0.41}
\definecolor{orange}{rgb}{0.85, 0.55, 0.13}
\definecolor{silver}{rgb}{0.69, 0.67, 0.66}
\definecolor{rust}{rgb}{0.72, 0.26, 0.06}
\definecolor{purp}{RGB}{68, 14, 156}
\definecolor{revisionblue}{RGB}{0, 82, 155}

\colorlet{lightrust}{rust!50!white}
\colorlet{lightorange}{orange!25!white}
\colorlet{lightlightblue}{lightblue}
\colorlet{lightsilver}{silver!30!white}
\colorlet{darkorange}{orange!75!black}
\colorlet{darksilver}{silver!65!black}
\colorlet{darklightblue}{lightblue!65!black}
\colorlet{darkrust}{rust!85!black}
\colorlet{darkseagreen}{seagreen!85!black}

\usepackage{hyperref}
\hypersetup{
  colorlinks=true,
}

\usepackage[labelfont=bf,font=small]{caption}

\usepackage[nameinlink]{cleveref}
\crefname{equation}{}{}
\def\appendixname{}
\crefname{appendix}{}{}

\usepackage[parfill]{parskip}

\usepackage{bm}

\newcommand{\overbar}[1]{\mkern 1.5mu\overline{\mkern-1.5mu#1\mkern-1.5mu}\mkern 1.5mu}

\usepackage{subcaption}
\usepackage{tikz}
\usetikzlibrary{external}
\usetikzlibrary{spy}

\usepackage{bm}

\usepackage{amsmath,amsfonts,amssymb,amsthm}
\usepackage{bbm}
\usepackage{pgfplotstable}
\usepackage{pgfplots}
\pgfplotsset{compat=1.18}
\usepgfplotslibrary{groupplots}
\usetikzlibrary{arrows}
\usetikzlibrary{shapes,snakes}
\usetikzlibrary{decorations.text}
\usepgfplotslibrary{fillbetween}

\definecolor{RYB1}{rgb}{0.72, 0.26, 0.06}
\definecolor{RYB2}{rgb}{0.18, 0.42, 0.41}
\definecolor{RYB3}{rgb}{0.63, 0.74, 0.78}
\definecolor{RYB4}{RGB}{251,220,127}
\definecolor{RYB5}{rgb}{0.69, 0.67, 0.66}
\definecolor{RYB6}{rgb}{0.85, 0.55, 0.13}

\definecolor{RYB7}{RGB}{128, 177, 211}

\pgfplotscreateplotcyclelist{newcolors}{
{RYB1,every mark/.append style={fill=RYB1,mark size={2}},mark=*},
{RYB2,every mark/.append style={fill=RYB2},mark=square*},
{RYB3,every mark/.append style={fill=RYB3,mark size={3}},mark=triangle*},
{RYB4,every mark/.append style={fill=RYB4,mark size={3}},mark=diamond*},
{RYB5,every mark/.append style={fill=RYB5,mark size={3}},mark=pentagon*},
{RYB6,every mark/.append style={fill=RYB6,mark size={4}},mark=10-pointed star},
{RYB7,every mark/.append style={fill=RYB7},mark=*},
}

\pgfplotsset{
    standard/.style={
    compat=1.18,
    enlarge x limits=0.00,
    enlarge y limits=0.00,
	cycle list name=newcolors,
    legend cell align={left},
    every axis/.append style={font=\small},
	every legend/.append style={font=\small},
	every node/.append style={font=\small},	
	}
}

\makeatletter
\@namedef{panellabel@at@south west}{0.015,0.015}
\@namedef{panellabel@at@north west}{0.015,0.985}
\@namedef{panellabel@at@south east}{0.985,0.015}
\@namedef{panellabel@at@north east}{0.985,0.985}
\newcommand{\panellabel}[2][south west]{%
    \@ifundefined{panellabel@at@#1}{%
        \PackageError{preamble}{\string\panellabel: unknown corner `#1'}%
            {Use one of: south west, north west, south east, north east.}%
    }{%
        \node[
            anchor=#1,
            inner sep=2.5pt,
            fill=white,
            fill opacity=0.7,
            text opacity=1,
            font=\small
        ] at (rel axis cs:\@nameuse{panellabel@at@#1}) {#2};%
    }%
}
\makeatother

\definecolor{lightblue}{rgb}{0.63, 0.74, 0.78}
\definecolor{seagreen}{rgb}{0.18, 0.42, 0.41}
\definecolor{orange}{rgb}{0.85, 0.55, 0.13}
\definecolor{silver}{rgb}{0.69, 0.67, 0.66}
\definecolor{rust}{rgb}{0.72, 0.26, 0.06}
\definecolor{purp}{RGB}{68, 14, 156}

\colorlet{lightsilver}{silver!30!white}
\colorlet{darkorange}{orange!75!black}
\colorlet{darksilver}{silver!65!black}
\colorlet{darklightblue}{lightblue!70!black}
\colorlet{darkrust}{rust!85!black}
\colorlet{darkseagreen}{seagreen!85!black}
\usepackage{xurl}
\providecommand{\doi}[1]{\href{https://doi.org/#1}{doi:#1}}

\begin{document}

\hypersetup{
    linkcolor=darkrust,
    citecolor=seagreen,
    urlcolor=darkrust,
    pdfauthor=author,
}

\begin{frontmatter}

\title{{\large\bfseries A sub-grid-scale model for polydisperse bubbly flows with heat and mass transfer}}

\author[1]{Anand Radhakrishnan}
\author[1,2,3]{Spencer H.\ Bryngelson}
\ead{shb@gatech.edu}

\address[1]{School of Computational Science \& Engineering, Georgia Institute of Technology, Atlanta, GA 30332, USA\vspace{-0.125cm}}
\address[2]{Daniel Guggenheim School of Aerospace Engineering, Georgia Institute of Technology, Atlanta, GA 30332, USA\vspace{-0.125cm}}
\address[3]{Woodruff School of Mechanical Engineering, Georgia Institute of Technology, Atlanta, GA 30332, USA}

\date{}

\begin{abstract}
    Ensemble-averaged models of polydisperse bubbly flows require statistics of the evolving bubble population.
    Prior quadrature-based moment formulations close bubble pressure with a polytropic relation that omits heat and mass transfer at the bubble wall.
    We formulate constant-transfer equations for bubble pressure and vapor mass within a conditional hyperbolic quadrature method.
    Second-order conditional inversion produces four joint radius--radial-velocity nodes per equilibrium-radius bin.
    Bubble pressure and vapor mass are advanced at each node.
    The node values close the ensemble-averaged flow equations without adding mixed pressure or vapor-mass moments to the transported moment set.
    The model is implemented in MFC.
    Monte Carlo calculations verify the evolution of quadrature nodes and the mean bubble variables for a harmonically forced population.
    Bubble-screen calculations quantify closure error as the equilibrium radius is discretized and the initial distributions vary.
    The constant-transfer calculation does not exhibit the high-frequency pressure oscillations observed with the polytropic closure under the conditions considered.
    3D bubble-screen calculations give a 1.5\% relative root-mean-square error between the Euler--Euler center pressure and the mean of 40 volume-averaged Euler--Lagrange realizations.
\end{abstract}


\end{frontmatter}

\blfootnote{The source code, example cases, random seeds, and detailed README have been archived on Zenodo as version 1.0.0 and are available at \doi{10.5281/zenodo.21590440}. The archived release corresponds to Git commit \texttt{c7d8fe135746}, \url{https://github.com/anandrdbz/MFC/tree/c7d8fe135746558bfb7c9d6f5103711d55b0e300}.}

\section{Introduction}

A wide range of length and time scales of the dispersed phase imposes strict restrictions on the problem size of resolved simulations.
Sub-grid-scale models can alleviate these high computational costs by modeling the dynamics of the dispersed phase through a statistical framework.
While each particle of the dispersed phase can be tracked using an Euler--Lagrange simulation~\citep{maeda18,vaca2026hardware}, performing these simulations is computationally expensive owing to poor scaling.
We therefore treat the dispersed phase in an Eulerian frame.
\citet{colonius08} showed that, under impulsive forcing, the moments of dilute bubble populations rapidly approach a statistical equilibrium and may be computed efficiently through period-averaged bubble dynamics, highlighting the benefits of reduced-order statistical representations.
The method of classes (MOC) employs ensemble phase averaging~\citep{zhang94, bryngelson19} to couple the dispersed and resolved phases.
The polydisperse bubble population is initialized using a log-normal distribution for the equilibrium radii.
The bubble dynamic variables are then modeled as a weighted sum of each bin of the log-normal distribution, with the weights obtained using Simpson's rule.
While the method of classes is an inexpensive alternative to Lagrangian methods, it employs only a stochastic framework for the equilibrium radius ($R_0$), with the internal variables fixed at each $R_0$.
Accurate representation of complex bubble dynamics requires incorporating stochasticity in the relevant internal bubble variables at each $R_0$.

The internal bubble variables at each $R_0$ are modeled using the population balance equation (PBE).
PBEs model the probability density of the dispersed phase~\citep{vanni00}.
The PBE uses the bubble radius and bubble velocity as the relevant internal variables and is solved via the method of moments (MOM)~\citep{mcgraw97}.
The MOM uses quadrature nodes to approximate unclosed high-order moments arising in the coupled flow equations~\citep{mcgraw97, Fox2018}.
PBE-based models can also accurately represent soot formation in the combustion process~\citep{Mueller2009}, aerosol sprays~\citep{sibra2017}, and more.
Here, a PBE-based method is used for its ability to represent the dynamics of the polydisperse bubbly phase~\citep{fox2008}.
Quadrature-based moment methods (QBMM) have emerged as an efficient Eulerian framework for representing polydisperse dispersed-phase distributions and have been successfully applied to bubbly flows and multivariate turbulent gas--liquid systems~\citep{heylmun19,Buffo2013}.
Here, we couple the constant-transfer equations for bubble pressure and vapor mass to the CHyQMOM quadrature representation~\citep{Fox2018}.
These variables are advanced at the joint radius--radial-velocity abscissae, and their node values supply the ensemble-averaged terms required by the compressible carrier-phase equations.
The sub-grid-scale model is coupled to the flow equations in the high-order, compressible flow solver MFC~\citep{wilfong2026mfc}.

Previous PBE-based models have considered the relevant internal variables as the bubble radius, bubble velocity, and equilibrium radius ($R$, $\dot{R}$, $R_0$).
The bubble radii and velocities of the polydisperse bubble population are initialized using a Gaussian distribution with means $R_0$ and $0$, respectively.
A polytropic assumption models the bubble pressure $p_b$ as a function of $R$ and $R_0$.
However, modeling complex bubble dynamics requires more sophisticated models, such as the constant-transfer model for bubble pressure~\citep{preston07}, to accurately represent spherical bubble dynamics.
Such models expand the set of relevant internal variables to include the bubble pressure $p_b$ and the vapor mass $m_v$.

The model extends \citet{bryngelson2023conditional}, in which bubble pressure is closed through a polytropic relation involving $R$ and $R_0$.
We instead advance $p_b$ and $m_v$ from the constant-transfer heat- and mass-transfer equations at the existing $(R,\Rdot)$ quadrature nodes for each equilibrium-radius bin.
The resulting node values close the ensemble-averaged terms without adding mixed moments in $p_b$ and $m_v$ to the transported moment set.

We implement the model in MFC~\citep{Bryngelson19_CPC,radhakrishnan23} and compare the node evolution and mean bubble variables with Monte~Carlo calculations for a harmonically forced population.
The bubble-screen calculations examine how the equilibrium-radius discretization and initial distributions affect closure error.
We compare the resulting pressure response with the polytropic closure and, in three dimensions, with the mean of 40 volume-averaged Euler--Lagrange realizations~\citep{vaca2026hardware}.

\section{Model formulation}\label{s:mfm}

\subsection{Compressible flow equations}

The governing equations for the liquid phase follow from the mixture-averaged compressible flow equations as
\begin{align}
    \frac{ \partial \rho }{\partial t} + \nabla \cdot ( {\rho \bu} ) &= 0, \nonumber \\
    \frac{ \partial {\rho \bu} }{\partial t} + \nabla \cdot ( \rho \bu \bu + p \bI ) &= 0, \label{e:euler}  \\
    \frac{ \partial {E} }{\partial t} + \nabla \cdot \left[({E} + p) \bu\right] &= 0, \nonumber
\end{align}
where $\rho$, $\bu$, $p$, and $E$ are the density, velocity vector, pressure, and total energy.
The density of the dispersed phase is considered negligible compared to the liquid phase.
The velocity of the liquid and dispersed phases is assumed to be the same, thus constituting a no-slip condition.
The contribution of the dispersed phase to the flow equations in \eqref{e:euler} is characterized through ensemble phase averaging.

\subsection{Ensemble phase averaging}

We use ensemble phase-averaged equations from prior work~\citep{zhang94,ando11,bryngelson19}.
The void fraction of the dispersed phase $\alpha_b$ is assumed to be negligible compared to the liquid phase ($\alpha_l$) under dilute assumptions.
The bubbles are represented statistically via random variables $R$, $\Rdot$, and $R_0$ corresponding to the instantaneous bubble radius, time derivative, and equilibrium bubble radius, presented in the next section.
The mixture-averaged pressure field is computed as
\begin{gather}
    p(\bx,t) = (1-\alpha) p_l +
    \alpha  \left(
        \frac{\overbar{R^3 p_{bw} }}{\overbar{R^3}} - \rho \frac{ \overbar{ R^3 \dot{R}^2 }}{ \overbar{R^3} }
    \right),
    \label{e:pressure}
\end{gather}
where $p_{bw}$ is the associated bubble wall pressure and $p_l(\bx,t)$ is the liquid pressure according to the stiffened-gas equation of state~\citep{menikoff89}:
\begin{gather}
    p_l + \gamma_l \pi_{\infty,l} =
    \frac{\gamma_l - 1}{1-\alpha} \left( E - \frac{1}{2} \rho \bu^2 \right).
    \label{e:SEOS}
\end{gather}
where $\gamma_l$ is the specific heat ratio of the liquid and $\pi_{\infty,l}$ is the liquid stiffness.
The bubble number density per unit volume $n(\bx,t)$ is conserved as
\begin{gather}
    \frac{\partial n }{\partial t } + \nabla \cdot ( n \bu ) = 0.
    \label{e:consn}
\end{gather}
For the spherical bubbles considered here, $n$ is related to the void fraction $\alpha$ via
\begin{gather}
    \alpha(\bx,t) = \frac{4}{3} \pi { \overbar{R^3} } n(\bx,t),
    \label{e:ndf}
\end{gather}
and thus the void fraction $\alpha(\bx,t)$ transports as
\begin{gather}
    \frac{\partial \alpha }{\partial t } +  \nabla \cdot (\alpha \bu) = \alpha \nabla \cdot(u) +
    3 \alpha \frac{ \overbar{R^2 \dot{R} }}{ \overbar{R^3} },
    \label{e:alpha}
\end{gather}
where the right-hand side represents the change in void fraction due to bubble growth and collapse.

The over-barred terms appearing in~\eqref{e:pressure}, \eqref{e:ndf}, and \eqref{e:alpha},
\begin{gather}
    \overbar{ R^3 \dot{R}^2},  \;
    \overbar{R^3}, \;
    \overbar{R^2 \dot{R} }, \; \text{and }
    \overbar{R^3 p_{bw}}
    \label{e:fullmoments}
\end{gather}
denote average quantities of the bubble dispersion.
The moments $\mom_{lmn}$ are averaged using a probability distribution function (PDF) $f(R,\Rdot, R_0)$ given as
\begin{align}
    \mom_{lmn} = \overbar{R^l \Rdot^m R_0^n}
        &= \int_\Omega R^l \Rdot^m R_0^n f(R,\Rdot,R_0) \, \ddd \dd R_0.
    \label{e:raw}
\end{align}

For the method of classes, the instantaneous bubble radius and velocity are fixed, reducing the PDF $f(R, \Rdot, R_0) = f(R_0)$, where $f(R_0)$ is log-normal with mean $\mu_{R_0} = 1$ and standard deviation $\sigma_{R_0}$.
$f(R_0)$ is fixed since $R_0$ is not a function of time.
The polydispersity of the bubble population sets the number of bins $n_b$, and \eqref{e:raw} is integrated using Simpson's rule.

\begin{figure*}
    \centering
    \includegraphics[scale=1]{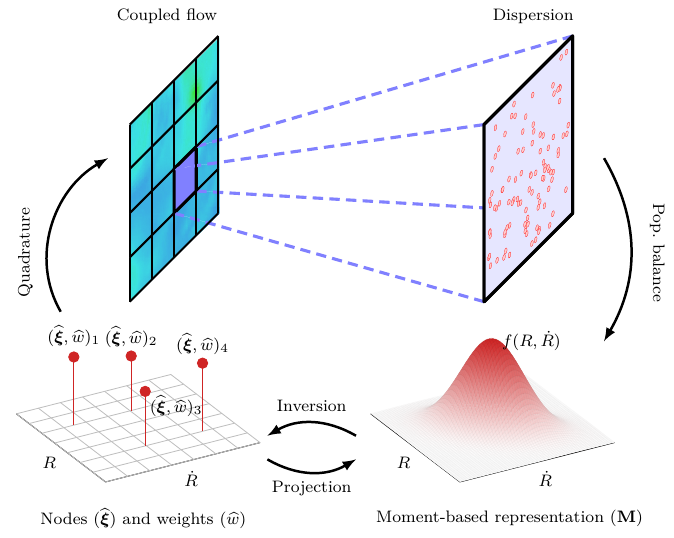}
    \caption{
        Illustration of our moment method (adapted from \citet{bryngelson2020qbmmlib} with permission)
    }
    \label{f:schematic}
\end{figure*}

\subsection{Population balance formulation and inversion}
We provide the formulation for the method of moments following that of \citet{bryngelson2023conditional}.
To account for stochasticity in $R$ and $\Rdot$, a population balance equation (PBE) is required for the evolution of the PDF.
The conditional probability density function $f(R,\dot{R}|R_0)$ without coalescence or breakup effects is governed by the following PBE~\citep{vanni00}.
\begin{gather}
    \frac{ \partial f}{\partial t} +
    \frac{\partial}{\partial R} (f \Rdot ) +
    \frac{\partial}{\partial \Rdot} (f \ddot{R} ) = 0
    \label{e:master}
\end{gather}
\Cref{f:schematic} illustrates the PBE-based approach.

The evolution of the raw moments $\vbmom$ is then obtained by integrating the PBE.
The set of transport equations for $\vbmom$ is
\begin{gather}
    \frac{\partial n \vbmom}{\partial t} + \nabla \cdot (n \vbmom \bu) =
    n {\dot\vbmom} =
    n \vbg
    \label{e:transport}
\end{gather}
where
\begin{gather}\label{e:rhs}
    g_{lmn} = l \mom_{l-1,m+1,n} + m \iiint_\Omega \Rddot  R^l \Rdot^{m-1} R_0^n f(R,\Rdot,R_0) \, \ddd \dd R_0
\end{gather}
and $\Omega = \Omega_R \times \Omega_{\Rdot} \times \Omega_{R_0} = (0,\infty) \times (-\infty,\infty) \times (0,\infty)$~\citep{bryngelson2020qbmmlib}.
The integral is computed via quadrature nodes obtained by a conditional hyperbolic inversion procedure~\citep{Fox2018}.
The bubble dynamics $\Ddot{R}$ is described in \cref{s:bubble_dyn}.
The PDF $f(R,\Rdot,R_0)$ is then related to the conditional PDF via
\begin{gather}
    f(R,\Rdot,R_0) = f(R,\Rdot|R_0) f(R_0),
\end{gather}
and the raw moments are obtained as
\begin{gather}
    \mom_{lmn} \equiv \int_{\Omega_{R_0}} f(R_0) R_0^n \mom_{lm}(R_0) \, \dd R_0
        \approx \sum_{i=1}^{n_{b}} w_i \hR_{0,i}^n \, \mom_{lm}(\hR_{0,i}),
    \label{e:full}
\end{gather}
where $n_{b}$ represents the polydispersity of the bubble population.
The quadrature rule weights and nodes for $R_0$ are obtained through the same procedure as the method of classes.
The $R_0$-conditioned moments $\mom_{lm}$ are then given by
\begin{gather}
    \mom_{lm}(\hR_{0,i}) \equiv \iint_{\Omega_{R,\Rdot}} f(R,\Rdot|\hR_{0,i}) R^l \Rdot^m \, \dd R \dd \Rdot
        \approx \sum_{j=1}^{N_R} \sum_{k=1}^{N_{\Rdot}} \left[ \hw_{j,k} \hR_j^l \hRdot_k^m \right]_{\hR_{0,i}}.
    \label{e:conditioned}
\end{gather}
The moment indices comprising the moment set of \eqref{e:conditioned} are associated with the conditional quadrature moment method used to invert those moments~\citep{yuan11,patel19}.
The CHyQMOM algorithm is described in~\citet{Fox2018}.
The total moments~\eqref{e:full} are obtained by substituting~\eqref{e:conditioned} into \eqref{e:full}:
\begin{gather}
    \mom_{lmn} =
        \sum_{i=1}^{n_{b}}
        w_i \hR_{0,i}^n
        \sum_{j=1}^{N_R} \sum_{k=1}^{N_{\Rdot}} \left[ \hw_{j,k} \hR_j^l \hRdot_k^m \right]_{\hR_{0,i}}.
\end{gather}
These moments evaluate the right-hand side of the moment evolution equation \eqref{e:transport} and the ensemble-phase averaged terms \eqref{e:fullmoments}.

\subsection{Bubble dynamics model}\label{s:bubble_dyn}

We model the bubble's radial dynamics with the Keller--Miksis equation, the compressible form of the Rayleigh--Plesset equation~\citep{plesset77}.
It captures complex bubbly-flow phenomena.
This model assumes spherical bubbles filled with a gas-vapor mixture.
Under these assumptions, the bubble radius is governed by a Rayleigh--Plesset-like equation
\begin{gather}\label{e:rpe}
    R \ddot{R} \left(1-\frac{\Rdot}{c_l}\right) + \frac{3}{2} \Rdot^2\left(1-\frac{\Rdot}{3c_l}\right)
    = \frac{p_{bw} - p_l}{\rho_l}\left(1+\frac{\Rdot}{c_l}\right) + \frac{R\,\dot{p}_{bw}}{\rho_l c_l},
\end{gather}
which is dimensionless via the reference bubble size $R_{0,\mathrm{ref}}$, liquid pressure $p_0$, and density $\rho_0$.
Since the bubble radius in our cases is much smaller than the imposed acoustic wavelength, the radiation term excludes the temporal variation in the liquid pressure $p_l$.
This reduces to the Rayleigh--Plesset equation under the assumption $\Rdot \ll c_l$, where $c_l$ represents the liquid speed of sound.
Here, $\rho_l$ is the liquid density, and the bubble wall pressure $p_{bw}$ is given as
\begin{gather}\label{e:pbw}
    p_{bw} = p_b - \frac{4\mu_l\Rdot}{R} - \frac{2\sigma_l}{R}
\end{gather}
with $p_b$ being the internal bubble pressure and $\mu_l$ and $\sigma_l$ being the dynamic viscosity and surface tension of the liquid phase, respectively.
The evolution of $p_b$ follows~\citep{preston07}
\begin{gather}
\begin{split}
    \dot{m}_v &= \frac{4 \pi R^2 D \rho_{bw} }{1 - \chi_{vw}} \frac{\partial \chi_{vw}}{\partial r}\Bigr|_{r=R}, \\
    \dot{p}_b &= \frac{3 \gamma_b}{R} \left(- \Dot{R} p_b + R_v T_{bw} \frac{\dot{m}_v}{4\pi R^2} + \frac{\gamma_b - 1}{\gamma_b} k_{bw} \frac{\partial T}{\partial r}\Bigr|_{r=R} \right),
    \label{e:pbmv}
\end{split}
\end{gather}
where $\chi_{vw}$, $k_{bw}$, $\rho_{bw}$, and $T_{bw}$ are the vapor mass fraction, thermal conductivity, density, and temperature at the bubble wall.
$D$ is the binary diffusion coefficient, $R_v$ is the gas constant, and $\gamma_b$ is the specific heat ratio.
The bubble wall closures in \eqref{e:pbmv} are defined in \citet{preston07}.

\subsection{Integrating constant-transfer model with the method of moments}
The conditional moments $\mu_{lm}(\hR_{0,i})$ are limited to second order for our implementation, such that $l + m = 2$.
Closing the moments, therefore, requires the first- and second-order moments of the initial distribution.
A joint normal distribution supplies them, with means $\mu_R$ and $\mu_{\Rdot}$, standard deviations $\sigma_R$ and $\sigma_{\Rdot}$, and a correlation coefficient $\rho_{R\Rdot}$ at each bin $\hR_{0,i}$.
The conditional inversion procedure outlined in \citet{Fox2018} thus uses $4$ nodes in the joint phase space of $R$ and $\Rdot$, visualized in \cref{f:schematic}.
A log-normal PDF is used for $R_0$ to represent the polydispersity in the flow with mean $\mu_{R_0} = 1$ and standard deviation $\sigma_{R_0}$.
The individual bins $\hR_{0,i}$ are subsequently calculated using Simpson's rule.
Evaluating $\overbar{R^3 p_{bw}}$ in \eqref{e:pressure} requires the statistics of the bubble pressure $p_b$.
Without polytropic assumptions, this requires including additional internal variables in the PBE.
To reduce the computational cost, the ODEs for $p_b$ and $m_v$ in \eqref{e:pbmv} are evolved at the quadrature nodes obtained by the moment inversion procedure for $R$ and $\Rdot$.
The moment set $\mu_{lm}(\hR_{0,i})$ is initialized as follows
\begin{gather}\label{e:mom_init}
\begin{gathered}
    \mom_{0,0}(\hat{R}_{0,i}) = 1, \qquad
    \mom_{1,0}(\hat{R}_{0,i}) = \hat{R}_{0,i}\mu_R, \qquad
    \mom_{0,1}(\hat{R}_{0,i}) = \sqrt{\frac{p_0}{\rho_0}}\,\mu_{\Rdot}, \\
    \mom_{2,0}(\hat{R}_{0,i}) = \hat{R}_{0,i}^2\left(\mu_R^2 + \sigma_R^2\right), \qquad
    \mom_{1,1}(\hat{R}_{0,i}) = \hat{R}_{0,i}\sqrt{\frac{p_0}{\rho_0}}\left(\mu_{R}\mu_{\Rdot} + \rho_{R\Rdot} \sigma_R \sigma_{\Rdot}\right), \\
    \mom_{0,2}(\hat{R}_{0,i}) = \frac{p_0}{\rho_0}\left(\mu_{\Rdot}^2 + \sigma_{\Rdot}^2\right)
\end{gathered}
\end{gather}
where $p_0$ and $\rho_0$ are the reference pressure and density at ambient conditions.
The bubble radius abscissa $R_{j,i}$ is calculated using the CHyQMOM procedure~\citep{Fox2018} at all $4$ quadrature nodes as follows
\begin{gather}
    \overbar{R}_{\pm,i} = \mom_{1,0,i} \pm \sqrt{\mom_{2,0,i} - \mom_{1,0,i}^2}, \qquad
    R_{1,i} = R_{2,i} = \overbar{R}_{-,i}, \qquad
    R_{3,i} = R_{4,i} = \overbar{R}_{+,i}
    \label{e:R_qbmm}
\end{gather}
The conditional velocity abscissae $\Rdot_{j,i}$ at the quadrature nodes are subsequently calculated as
\begin{gather}
\begin{gathered}
    \overbar{\Rdot}_{\pm,i} = \mom_{0,1,i} \pm \frac{\mom_{1,1,i} - \mom_{1,0,i}\mom_{0,1,i}}{\sqrt{\mom_{2,0,i} - \mom_{1,0,i}^2}}, \qquad
    s_{i} = \sqrt{\mom_{0,2,i} - \mom_{0,1,i}^2 - \frac{(\mom_{1,1,i} - \mom_{1,0,i}\mom_{0,1,i})^2}{\mom_{2,0,i} - \mom_{1,0,i}^2}}, \\
    \Rdot_{1,i} = \overbar{\Rdot}_{-,i} - s_{i}, \qquad
    \Rdot_{2,i} = \overbar{\Rdot}_{-,i} + s_{i}, \qquad
    \Rdot_{3,i} = \overbar{\Rdot}_{+,i} - s_{i}, \qquad
    \Rdot_{4,i} = \overbar{\Rdot}_{+,i} + s_{i}
\end{gathered}
    \label{e:Rdot_qbmm}
\end{gather}
The bubble radius $R_{j,i}$ is initialized at the $4$ quadrature nodes for all $\hat{R}_{0,i}$ bins as
\begin{gather}\label{e:quad_init}
    R_{1,i} = R_{2,i} = \hat{R}_{0,i}\left(\mu_R - \sigma_R\right), \qquad
    R_{3,i} = R_{4,i} = \hat{R}_{0,i}\left(\mu_R + \sigma_R\right)
\end{gather}
Parameters $\mu_R$ and $\sigma_R$ are normalized using the equilibrium radius $\hat{R}_{0,i}$ at each bin.
The bubble pressure $p_{b_{j,i}}$ and vapor mass $m_{v_{j,i}}$ are subsequently set at each of the quadrature nodes using the following isothermal assumptions owing to small errors in the ensuing bubble dynamics between the constant-transfer and isothermal models~\citep{preston07}.
\begin{gather}\label{e:quad_init_pbmv}
    m_{v_{j,i}} = m_{v_{0,i}} \frac{R_{j,i}^3}{\hat{R}_{0,i}^3}, \qquad
    p_{b_{j,i}} = p_{b_{0,i}} \frac{\hat{R}_{0,i}^3\left(m_{g_{0,i}} + m_{v_{j,i}}\right)}{R_{j,i}^3\left(m_{g_{0,i}} + m_{v_{0,i}}\right)}
\end{gather}
where $m_{g_{0,i}}$ and $m_{v_{0,i}}$ are the mass of gas and vapor at the equilibrium radius $\hat{R}_{0,i}$ evaluated using the ideal gas law.
This procedure implicitly assumes the probability densities of $p_b$ and $m_v$ to be delta functions centered at the quadrature nodes for $R$ and $\dot{R}$.
The bubble dynamic equations in \eqref{e:pbmv} are subsequently evaluated at the quadrature nodes for each of the $\hat{R}_0$ bins as follows
\begin{gather}
\begin{split}
    \dot{m}_{v_{j,i}} &= \frac{4 \pi R_{j,i}^2 D \rho_{bw}}{1 - \chi_{vw}} \frac{\partial \chi_{vw}}{\partial r}\Bigr|_{r=R_{j,i}}, \\
    \dot{p}_{b_{j,i}} &= \frac{3 \gamma_b}{R_{j,i}} \left(- \frac{\partial R_{j,i}}{\partial t} p_{b_{j,i}} + R_v T_{bw} \frac{\dot{m}_{v_{j,i}}}{4\pi R_{j,i}^2} + \frac{\gamma_b - 1}{\gamma_b} k_{bw} \frac{\partial T}{\partial r}\Bigr|_{r=R_{j,i}} \right)
    \label{e:pbmvqbmm}
\end{split}
\end{gather}
where $\frac{\partial R_{j,i}}{\partial t}$ denotes the evolution of the radial abscissae for all $4$ quadrature nodes as opposed to the conditional velocity abscissae $\Rdot_{j,i}$.
$\frac{\partial R_{j,i}}{\partial t}$ is used to evolve $p_b$ as the equality to conditional velocity holds only for the ensemble average and not for individual quadrature nodes.
$\frac{\partial R_{j,i}}{\partial t}$ is calculated using the hyperbolic inversion procedure in \citet{Fox2018} as follows
\begin{gather}
\begin{split}
    \frac{\partial R_{1,i}}{\partial t} = \frac{\partial R_{2,i}}{\partial t} &= \frac{\partial \mom_{1,0,i}}{\partial t} - \frac{\dfrac{\partial \mom_{2,0,i}}{\partial t} - 2\mom_{1,0,i} \dfrac{\partial \mom_{1,0,i}}{\partial t}}{2\sqrt{\mom_{2,0,i} - \mom_{1,0,i}^2}}, \\
    \frac{\partial R_{3,i}}{\partial t} = \frac{\partial R_{4,i}}{\partial t} &= \frac{\partial \mom_{1,0,i}}{\partial t} + \frac{\dfrac{\partial \mom_{2,0,i}}{\partial t} - 2\mom_{1,0,i} \dfrac{\partial \mom_{1,0,i}}{\partial t}}{2\sqrt{\mom_{2,0,i} - \mom_{1,0,i}^2}},
    \label{e:drdt}
\end{split}
\end{gather}
The temporal evolution of the moments $\mom_{1,0,i}$ and $\mom_{2,0,i}$ is then formulated in terms of the conservative transport equations~\eqref{e:transport} as follows
\begin{gather}
    \frac{\partial \mom_{1,0,i}}{\partial t} = \frac{1}{n} \left(\frac{\partial \left(n \mom_{1,0,i}\right)}{\partial t} -  \mom_{1,0,i} \frac{\partial n}{\partial t}\right), \qquad
    \frac{\partial \mom_{2,0,i}}{\partial t} = \frac{1}{n} \left(\frac{\partial \left(n \mom_{2,0,i}\right)}{\partial t} -  \mom_{2,0,i} \frac{\partial n}{\partial t}\right)
    \label{e:dmudt_prim}
\end{gather}
where $\partial (n\mom_{1,0,i})/\partial t$ and $\partial (n\mom_{2,0,i})/\partial t$ are calculated using \eqref{e:transport}.
$\partial n / \partial t$ in \eqref{e:dmudt_prim} is calculated using $\mom_{0,0,i} = 1$ as
\begin{gather}
    \frac{\partial n}{\partial t} + \nabla \cdot (n \bu) = 0,
    \label{e:dndt}
\end{gather}
recovering the number-density conservation law \eqref{e:consn}.
When $\mom_{2,0,i} - \mom_{1,0,i}^2 < 0$, the moments are not realizable.
The implementation floors the radius variance in the univariate inversion,
\begin{gather}
    \mom_{2,0,i} - \mom_{1,0,i}^2
    \leftarrow \max\left(\mom_{2,0,i} - \mom_{1,0,i}^2,\ \sigma_{\epsilon}\right),
    \label{e:limiter}
\end{gather}
where $\sigma_{\epsilon} = \num{e-16}$. The residual conditional velocity is also floored at $\sigma_{\epsilon}$.

\section{Numerical method}\label{s:numerics}

An interface-capturing finite-volume scheme evolves the governing equations for the liquid and dispersed phases.
The conservative variables $\vbq_c$ are evolved in the cell centers according to
\begin{gather}
    \frac{\partial\vbq_c}{\partial t} + \nabla \cdot \vbF = \vbs
    \label{e:governing}
\end{gather}
The source vector $\vbs$ includes the moment sources $n\vbg$ of \cref{e:transport} and the void-fraction growth/collapse source on the right-hand side of \cref{e:alpha}.
In the current MFC implementation, $p_{b_{j,i}}$ and $m_{v_{j,i}}$ are stored in separate node arrays rather than in $\vbq_c$ or $\vbs$, and no conservative transport is applied to them.
Their right-hand sides are evaluated using \eqref{e:pbmvqbmm} at every SSP--RK stage and updated with the same RK coefficients as the conservative state, without sub-cycling.
The flux-induced corrections for $p_b$ enter through \eqref{e:drdt}, \eqref{e:dmudt_prim}, and \eqref{e:dndt}, and are required by the temporal evolution term $\partial R_{j,i}/\partial t$ on the right-hand side of \eqref{e:pbmvqbmm}.
The evolution of $m_v$ requires no such correction, as its right-hand side in \eqref{e:pbmvqbmm} contains no temporal evolution term.
The conservative variables $\vbq_c$ are converted to their primitive forms $\vbq_p$.
The primitive variables are then reconstructed at the cell boundaries using a fifth-order WENO scheme~\citep{jiang1996efficient}.
The fluxes $\vbF$ are computed using an HLLC approximate Riemann solver~\citep{toro94}.
A third-order Runge--Kutta scheme advances the conservative variables.
Our implementation is available in MFC, an open-source CFD solver~\citep{wilfong2026mfc}.

\section{Model verification}

We examine the ability of the model to accurately represent bubble dynamics for a monodisperse air bubble suspended in water.
The liquid pressure $p_l$ is explicitly set to $p_l = p_0(1 + A \sin(\omega t))$ where $p_0$ is the reference pressure and $t$ is the simulation time.
This setup ignores the governing equations in \eqref{e:euler} and only evolves the moment transport equations~\eqref{e:transport} along with $p_b$ and $m_v$ arrays to verify the ability of the model to capture bubble dynamics accurately.
For our simulation, we set $A = 0.5$ and $\omega = \pi$.
This corresponds to an excitation frequency that is $1.53$ times the Minnaert frequency.
We non-dimensionalize lengths by the reference equilibrium radius $R_{0,\mathrm{ref}} = \qty{10}{\micro\metre}$.
The bubbles are initialized at equilibrium $\mu_R = R_0 = 1$ and $\mu_{\Rdot} = 0$, with standard deviations $\sigma_R = \sigma_{\Rdot} = 0.1$ and $\rho_{R\Rdot} = 0$.

The reference pressure and density of the liquid are set as $p_0 = \qty{101325}{\pascal}$ and $\rho_0 = \qty{1000}{\kilogram\per\cubic\metre}$.
$\gamma_b$ is set at $1.4$, with the speed of sound $c_l = \qty{1475}{\metre\per\second}$.
Surface tension, viscosity, and vapor pressure of the liquid are set at $\sigma_l = \qty{0.0752}{\newton\per\metre}$, $\mu_l = \qty{0.001}{\pascal\second}$, and $p_v = \qty{2388}{\pascal}$.
The thermodynamic parameters associated with the constant-transfer model are calculated using a water~vapor and air mixture, with the thermodynamic properties and bubble wall closures defined by \citet{preston07}.
The bubble dynamics is simplified to use the Rayleigh--Plesset equation, which is the incompressible version of \eqref{e:rpe}.
The time step $\Delta t$ is set to $0.0005\,t_0$, where $t_0 = R_{0,\mathrm{ref}} \sqrt{\rho_0/p_0}$ is the reference time.

\begin{figure*}
    \centering
    \includegraphics[scale=1]{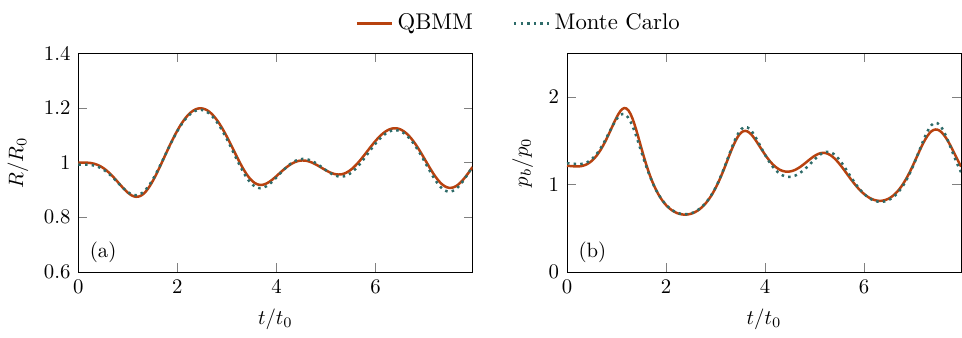}
    \caption{
        Comparison of the evolution of (a) the mean bubble radius and (b) the bubble pressure, for the constant-transfer QBMM and Monte Carlo simulations of a monodisperse bubble.
    }
    \label{f:err_time}
\end{figure*}

\begin{figure*}
    \centering
    \includegraphics[scale=1]{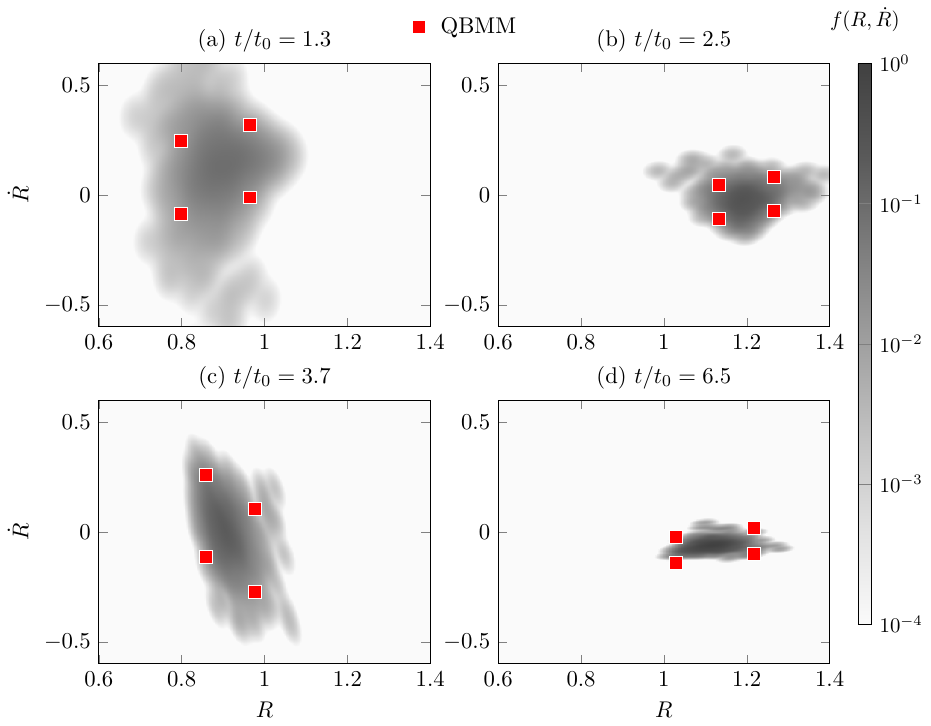}
    \caption{
        Evolution of probability density $f(R, \Rdot)$ obtained through Monte Carlo simulation with $N_s = 1000$ along with the quadrature nodes for the constant-transfer QBMM model at four different simulation times.
    }
    \label{f:quad_time}
\end{figure*}

To examine closure errors in the moment set, we compare results with Monte Carlo simulations sampled from a multivariate Gaussian distribution for $R$ and $\Rdot$, with the same means and standard deviations.
We generate $N_s = 1000$ individual samples for the Monte Carlo simulations using the class method and average them to obtain a surrogate for the true solution.
The choice of $N_s = 1000$ follows prior practice: \citet{charalampopoulos2022hybrid} report adequate accuracy at this sample count for similar Monte Carlo runs under the polytropic assumption.
We do not repeat that sample-count study here.
The RMS error $\|e\|_2$ for a particular bubble variable $v$ is calculated using

\begin{gather}
    \|e\|_2 = \frac{1}{\sqrt{N_t}}
    \sqrt{\sum_{i=0}^{N_t-1} \left(\frac{v^{(\mathrm{QBMM})}(t_i) - v^{(\mathrm{MC})}(t_i)}{v^{(\mathrm{MC})}(t_i)}\right)^{\!2}}
    \label{e:rms}
\end{gather}

where $v$ can be any of the relevant bubble variables $R$, $p_b$, and $m_v$, or the moments $\mu_{lm}$.
The RMS error $\|e\|_2$ for the bubble variables $R$, $p_b$, and $m_v$ for the constant-transfer model is compared against the polytropic model previously verified in \cref{tab:epsilon}.
The polytropic model carries no $p_b$ or $m_v$, so \cref{tab:epsilon} crosses out those entries.
The $1.4\times$ increase in errors in the constant-transfer model arises from the inexactness of the initial condition due to the ODE for $p_b$ and $m_v$ having no true solution.
This is verified by examining $\|e\|_2$ for the constant-transfer QBMM model by setting $k_{bw} = 0$ and using the infinite rate mass transfer limit, wherein the ODE for $p_b$ in \eqref{e:pbmv} changes as

\begin{gather}
    \dot{p}_b = -\frac{3 \gamma_b}{R} \Dot{R} \left(p_b - p_v\right)
    \label{e:pbpseudo}
\end{gather}
where $p_v$ is the vapor pressure.
This effectively yields a polytropic evolution for $p_b$ at the quadrature nodes, allowing an exact initial condition.
This case is labeled pseudo-polytropic in \cref{tab:epsilon} and agrees analytically with the polytropic closure in MFC, while explicitly evolving $p_b$ at the $(R, \Rdot)$ quadrature nodes using \eqref{e:pbpseudo}.
Error values for $R$ are identical for both the polytropic and pseudo-polytropic cases up to round-off, with $\|e\|_2 = \qty{0.58}{\percent}$.
This verifies that the increased error in $R$ of \qty{0.81}{\percent} for the constant-transfer model arises from an inexact initial condition and not the pinning of $p_b$ and $m_v$ at the quadrature nodes for $R$ and $\Rdot$.
The evolution of the bubble radius $R$ and bubble pressure $p_b$ for the constant-transfer model and Monte Carlo simulations is shown in \Cref{f:err_time}, with good agreement between the two.
\Cref{f:quad_time} verifies that the evolution of the quadrature nodes follows the probability density of the Monte Carlo samples at four different simulation times.

\begin{table}[h]
    \centering
    \begin{tabular}{cccc}
        \toprule
        $10^{2}\,\|e\|_2$ & Polytropic & Constant transfer & Pseudo polytropic \\
        \midrule
        $R$ & $0.58$ & $0.81$ & $0.58$ \\
        $p_b$ & \ding{55} & $2.62$ & \ding{55} \\
        $m_v$ & \ding{55} & $3.49$ & \ding{55} \\
        \bottomrule
    \end{tabular}
    \caption{
        RMS relative error $\|e\|_2$ between various QBMM models and Monte Carlo simulations with $N_s = 1000$
    }
    \label{tab:epsilon}
\end{table}

\section{Application to bubble screens}

We test our model by examining the bubble dynamics at the center of an acoustically excited polydisperse, dilute 1D bubble screen.
The bubble screen is initialized with a void fraction of $\alpha_b = \num{4e-5}$ and a log-normal PDF with mean radius $R_0 = 1$, non-dimensionalized by the reference bubble radius $R_{0,\mathrm{ref}} = \qty{10}{\micro\metre}$.
The bubble screen length is set to $L_b = 800R_0$ and is centered in a 1D domain of length $5L_b$, with $N_x = 1600$ grid points.
The thermodynamic parameters are reused from the previous section on model verification.
A one-way sinusoidal pressure pulse with peak amplitude $p_0$ excites the bubble screen, where $p_0$ is the ambient pressure.
The frequency of the sinusoidal pulse is set to \qty{300}{\kilo\hertz}, with the Minnaert frequency of the bubble population being \qty{326}{\kilo\hertz} with adiabatic assumptions ($\gamma_b = 1.4$) and negligible surface tension and viscous effects.
Neumann boundary conditions are used along the pulse direction.
The acoustic source is located $700 R_0$ from the center of the bubble screen with a support width spanning $15$ grid spacings.
The bubble population is polydisperse, with $n_b = \{9,21,31,51\}$ bins.
The closure errors are measured relative to $n_b = 91$, where the relative RMS error with $n_b = 51$ for a representative case is \qty{0.01}{\percent}, indicating a converged solution.
Previous work by \citet{bryngelson19} for the method of classes also suggests convergence in this range.
The standard deviation of the log-normal PDF is varied as $\sigma_{R_0} = \{0.1, 0.2, 0.3\}$.
The initial distribution for the moment set is a joint Gaussian distribution with means $\mu_R = 1$ and $\mu_{\Rdot} = 0$, with the standard deviations varied as $\sigma_R = \{0.1, 0.2, 0.3\}$ and $\sigma_{\Rdot} = \{0.1, 0.2, 0.3\}$.
We characterize the effects of varying $\sigma_R, \sigma_{\Rdot}, \sigma_{R_0}$ by plotting the pressure $p$ at the center of the bubble screen and evaluating the resulting closure errors at varying $n_b$.
The pressure is non-dimensionalized by the reference pressure $p_0$ and the time by $t_0 = R_{0,\mathrm{ref}}\sqrt{\rho_0/p_0}$.
Bubble dynamics is governed by the Keller--Miksis equations~\eqref{e:rpe}.
We then compare with the previously validated polytropic model~\citep{bryngelson2023conditional} to highlight differences in bubble dynamics.
Finally, the model is validated by comparison with a previously validated volume-averaged Euler--Lagrange model~\citep{vaca2026hardware} through 3D domain extension.
Periodic boundary conditions are applied along the transverse directions.
The CFL number for these simulations is set to $0.2$.

\begin{figure*}
    \centering
    \includegraphics[scale=1]{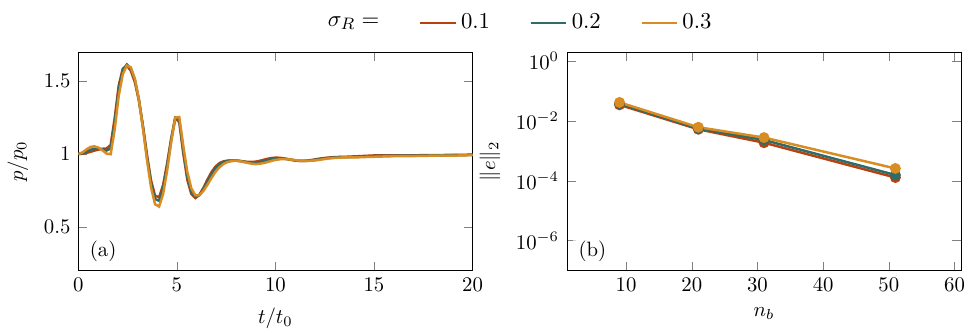}
    \caption{
        (a) Dimensionless pressure $p/p_0$ versus time $t/t_0$ at the center of the bubble screen at $n_b = 51$. (b) RMS relative error $\|e\|_2$ for the pressure for a varying number of bins $n_b$, with a reference solution at $n_b = 91$. Both are shown at varying values of $\sigma_R$ and fixed $\sigma_{\Rdot} = 0.2$ and $\sigma_{R_0} = 0.3$.
    }
    \label{f:sigR}
\end{figure*}

\begin{figure*}
    \centering
    \includegraphics[scale=1]{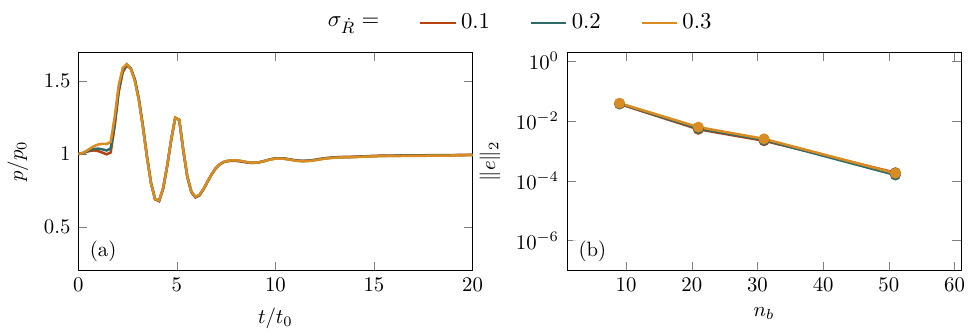}
    \caption{
        (a) Dimensionless pressure $p/p_0$ versus time $t/t_0$ at the center of the bubble screen at $n_b = 51$. (b) RMS relative error $\|e\|_2$ for the pressure for a varying number of bins $n_b$, with a reference solution at $n_b = 91$. Both are shown at varying values of $\sigma_{\Rdot}$ and fixed $\sigma_R = 0.2$ and $\sigma_{R_0} = 0.3$.
    }
    \label{f:sigV}
\end{figure*}

\begin{figure*}
    \centering
    \includegraphics[scale=1]{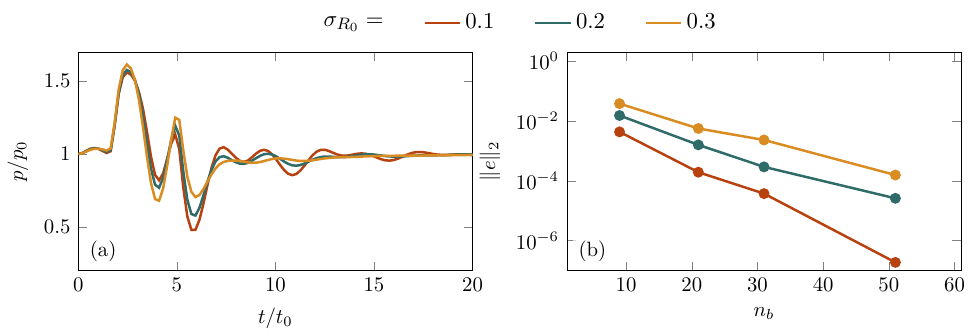}
    \caption{
        (a) Dimensionless pressure $p/p_0$ versus time $t/t_0$ at the center of the bubble screen at $n_b = 51$. (b) RMS relative error $\|e\|_2$ for the pressure for a varying number of bins $n_b$, with a reference solution at $n_b = 91$. Both are shown at varying values of $\sigma_{R_0}$ and fixed $\sigma_R = \sigma_{\Rdot} = 0.2$.
    }
    \label{f:sigR0}
\end{figure*}

\subsection{Varying $\sigma_R$}

We first consider the representative bubble screen problem by fixing $\sigma_{\Rdot} = 0.2$ and $\sigma_{R_0} = 0.3$ and varying $\sigma_R = \{0.1, 0.2, 0.3\}$.
The closure errors are plotted in \cref{f:sigR} at varying values of $n_b$ compared to a reference solution with $n_b = 91$.
The relative RMS error $\|e\|_2$ increases only slightly with $\sigma_R$.
This contrasts with the results in \citet{bryngelson2023conditional} for the polytropic model.
The pressure at the center of the bubble screen at $n_b = 51$ shows an absence of the higher frequency components associated with the natural frequency of the bubble present in \citet{bryngelson2023conditional}.
A slight decrease in the minimum radius during the collapse phase is observed with an increase in $\sigma_R$.

\subsection{Varying $\sigma_{\Rdot}$}

Next we fix $\sigma_{R} = 0.2$ and $\sigma_{R_0} = 0.3$ and vary $\sigma_{\Rdot} = \{0.1, 0.2, 0.3\}$.
The closure errors are plotted in \cref{f:sigV} at varying values of $n_b$ compared to a reference solution with $n_b = 91$.
The relative RMS error $\|e\|_2$ is essentially unchanged by $\sigma_{\Rdot}$.
This is consistent with the results presented in \citet{bryngelson2023conditional} for the polytropic model.
The pressure at the center of the bubble screen at $n_b = 51$ is accordingly insensitive to $\sigma_{\Rdot}$.
The effects of varying $\sigma_R$ on the resulting bubble dynamics are more pronounced when compared to those of varying $\sigma_{\Rdot}$.

\subsection{Varying $\sigma_{R_0}$}

Finally, we fix the initial condition for the moment set at $\sigma_R = \sigma_{\Rdot} = 0.2$ and vary $\sigma_{R_0} = \{0.1,0.2,0.3\}$ for the log-normal PDF.
The closure errors are plotted in \cref{f:sigR0} at varying values of $n_b$ compared to a reference solution with $n_b = 91$.
The relative RMS error $\|e\|_2$ increases with increasing $\sigma_{R_0}$.
The pressure at the center of the bubble screen shows a tendency toward persistent oscillations as $\sigma_{R_0}$ is decreased.
The high-frequency oscillations at larger values of $\sigma_{R_0}$ in \citet{bryngelson2023conditional} are absent with the constant-transfer model.
The increase in the relative error $\|e\|_2$ can be attributed to a broader distribution, leading to a larger truncation error from Simpson's rule.

\subsection{Comparison to polytropic model}

\begin{figure*}
    \centering
    \includegraphics[scale=1]{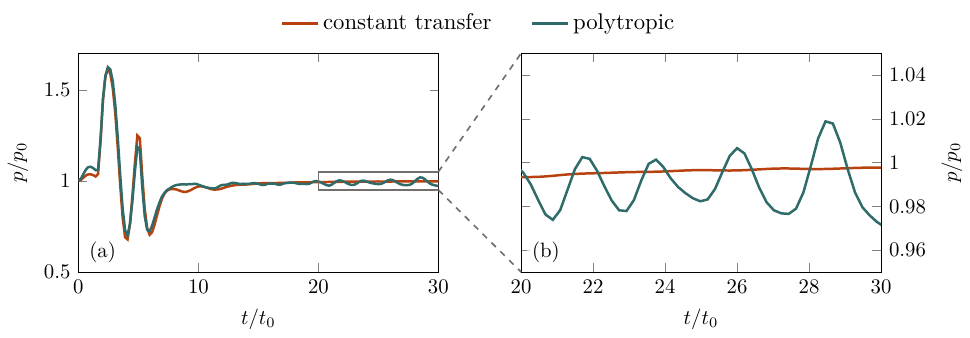}
    \caption{
        (a) Dimensionless pressure $p/p_0$ versus time $t/t_0$ at the center of the bubble screen for the constant-transfer model and the polytropic model at $\sigma_R = \sigma_{\Rdot} = 0.2$, $\sigma_{R_0} = 0.3$, and $n_b = 51$. (b) The outlined window of (a), $20 \leq t/t_0 \leq 30$, at which point the pressure has equilibrated, and the high-frequency oscillations of the polytropic closure are clearest.
    }
    \label{f:poly}
\end{figure*}
We compare the bubble dynamics at the center of the bubble screen for the constant-transfer model with those of the previously validated polytropic model.
The initial condition for the moment set is set at $\sigma_R = \sigma_{\Rdot} = 0.2$ and $\sigma_{R_0} = 0.3$ with $n_b = 51$.
The results are displayed in \cref{f:poly}, which compares the pressure at the center of the bubble screen.
While the effects of the large-wavelength input wave are similar in both models, the constant-transfer model suppresses the small-wavelength oscillations associated with the bubble screen's natural frequencies.
The simulation is run to $t/t_0 = 30$ to demonstrate the differences between the models clearly, and panel (b) of \cref{f:poly} shows a zoomed-in view of the pressure profiles for $t/t_0 > 20$.
As the pressure equilibrates with the passage of the sinusoidal wave, this difference can be observed more clearly for larger $t/t_0 > 20$.
This is consistent with the absence of the high-frequency components with increasing $\sigma_R$ and $\sigma_{R_0}$ observed in the previous section as compared to \citet{bryngelson2023conditional}.

\subsection{Comparison to Euler--Lagrange model}

\begin{figure*}
    \centering
    \includegraphics[scale=1]{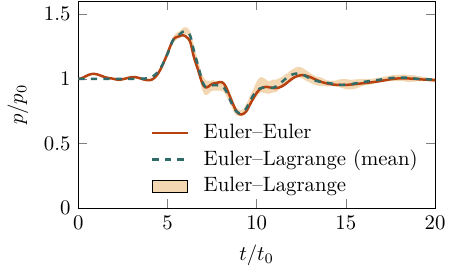}
    \caption{
        Dimensionless pressure $p/p_0$ versus time $t/t_0$ at the center of the 3D bubble screen for the constant-transfer Euler--Euler model, compared with the mean of the volume-averaged Euler--Lagrange model across $40$ independent simulations, with a band delineating the minimum and maximum values across those simulations, at $\sigma_R = \sigma_{\Rdot} = 0.2$, $\sigma_{R_0} = 0.3$, and $n_b = 51$.
    }
    \label{f:EL}
\end{figure*}
Finally, we validate our constant-transfer model with the previously validated Euler--Lagrange model~\citep{vaca2026hardware}.
The bubble screen length for this experiment is $L_b = 500R_0$, with the domain length $L_x = 8L_b$.
To perform this comparison, we extend the bubble screen in 3D with $L_y = L_z = 5L_b/8$ and apply periodic boundary conditions along the transverse directions.
The grid resolution is coarsened to $(N_x, N_y, N_z) = 400 \times 50 \times 50$.
The coarsening is done to ensure that the grid resolution is sufficiently low so that the maximum radius of the Lagrangian bubbles is less than the grid resolution, while still resolving the sinusoidal input.
The Lagrangian bubbles are placed uniformly within the bubble screen dimension $L_b$.
We set $\sigma_{R_0} = 0.3$ and sample the radii of the Lagrangian bubbles from the log-normal PDF to satisfy $\alpha_b = \num{4e-5}$ within the bubble screen.
We compute the mean of $N_{\text{sim}} = 40$ independent instances of the Euler--Lagrange simulation to sufficiently reduce sampling error as per \citet{bryngelson19, vaca2026hardware}.
The Euler--Euler simulation is conducted with $\sigma_R = \sigma_{\Rdot} = 0.2$ and $n_b = 51$.
The results are displayed in \cref{f:EL}, which compares the pressure at the center of the bubble screen for the Euler--Euler and the mean of the Euler--Lagrange simulations.
The RMS relative error between these simulations is computed to be $ \|e\|_2 = 0.015$, indicating excellent agreement.
The Euler--Euler curve is not pointwise contained within the minimum--maximum envelope of the individual Euler--Lagrange results during the early driven and collapse phases; its maximum departure from the envelope is approximately \qty{4}{\percent} of $p_0$.

\section{Conclusion}\label{s:conclusions}

We present an Euler--Euler stochastic sub-grid-scale model that uses the method of moments.
The population balance equation carries stochasticity in the dispersed-phase internal variables.
A conditional hyperbolic inversion procedure advances the moment transport equations and the ensemble-averaged terms.
A constant-transfer model represents complex bubble dynamics better than the previous polytropic model.
The PBE carries additional internal variables.
Delta functions at the quadrature nodes approximate the probability densities, reducing cost.
The model is incorporated into the high-order accurate compressible flow solver MFC.
We verified the model by comparing node evolution and bubble variables to Monte Carlo simulations for a monodisperse bubble.
The model was tested for an acoustically excited, polydisperse bubble-screen problem with a sinusoidal input pulse.
The constant-transfer model suppresses the high-frequency oscillations associated with the bubble screen's natural frequencies, compared to the polytropic model.
The initial distribution of the moment set is found to have a small effect on closure errors, with sensitivity increasing with broader equilibrium-radius distributions ($\sigma_{R_0}$).
The model is validated by comparing against the mean of previously validated volume-averaged Euler--Lagrange simulations.
The model shows excellent agreement on the bubble dynamics in the center of the bubble screen, with an RMS relative error of \qty{1.5}{\percent}.

The current limitation of the polydisperse method of moments is instability at large pressure ratios when paired with a small equilibrium radius.
This limitation persists even with the previous polytropic implementation, when the bubble radius rapidly approaches zero.
This currently causes stability issues in cases such as flow over an immersed boundary, where a strong shock wave passes through the domain, leading to non-realizable moments.
Future work would aim to resolve these stability issues by investigating limiters for non-realizable moments and adaptive sub-stepping for the dispersed phase.

\section*{CRediT authorship contribution statement}

\textbf{Anand Radhakrishnan:} Conceptualization, Methodology, Software, Validation, Formal analysis, Investigation, Data curation, Visualization, Writing --- original draft. \\
\textbf{Spencer H.\ Bryngelson:} Conceptualization, Methodology, Resources, Supervision, Funding acquisition, Writing --- review \& editing.

\section*{Declaration of competing interest}

The authors declare that they have no known competing financial interests or personal relationships that could have appeared to influence the work reported in this paper.

\section*{Acknowledgments}

SHB acknowledges support from the US Office of Naval Research under grant number N00014-22-1-2519.
This research used resources of the Oak Ridge Leadership Computing Facility at the Oak Ridge National Laboratory, which is supported by the Office of Science of the U.S.\ Department of Energy under Contract No.~DE-AC05-00OR22725 (allocation CFD154, PI~Bryngelson).
This work also used PSC~Bridges2 and NCSA~Delta and Delta~AI through allocation PHY210084 (PI~Bryngelson) from the Advanced Cyberinfrastructure Coordination Ecosystem: Services \& Support (ACCESS) program, which is supported by National Science Foundation grants \#2138259, \#2138286, \#2138307, \#2137603, and \#2138296.

\section*{Data availability}

The datasets generated during this study include simulation input files, example cases, random seeds, and the source code used to produce the reported results. These materials have been archived on Zenodo (version 1.0.0) and are publicly available at \doi{10.5281/zenodo.21590440}, corresponding to Git commit \texttt{c7d8fe135746}.

\bibliography{main.bib}
\end{document}